\documentclass[referee]{aa}
\usepackage{txfonts}
\usepackage{graphicx}
\usepackage{epstopdf}
\usepackage{longtable}
\usepackage{lscape}
\usepackage{natbib}
\bibpunct{(}{)}{;}{a}{}{,} 

\begin{document}

\title{Far-infrared study of K giants in the solar neighborhood:\\ Connection between Li enrichment and mass-loss}
\author{Yerra Bharat Kumar $^{1,2}$, B. Eswar Reddy$^{2}$, C. Muthumariappan$^{2}$, G. Zhao$^{1}$}

\institute{$^{1}$Key Laboratory of Optical Astronomy, National Astronomical Observatories, 
Chinese Academy of Sciences, Beijing 100012, China \\
$^{2}$Indian Institute of Astrophysics, Koramangala II Block, Bangalore 560034, India}
\date{Received,Accepted}

\begin{abstract}
{A small group of red giant branch (RGB) stars are known to have anomalous Li enhancement
whose origin is still not well understood. 
Some studies have proposed that the Li enhancement in RGB stars is
correlated to their far-IR excess, a result of mass loss.  
Studies to confirm such a correlation have a significant bearing 
on our understanding of the Galactic Li enhancement.  
}
{We searched for a correlation between the two anomalous properties of K giants: Li enhancement and IR excess
from an unbiased survey of a large sample of RGB stars.}
{A sample of 2000 low-mass K giants with accurate astrometry
from the Hipparcos catalog was chosen for which Li abundances have been determined from
low-resolution spectra. Far-infrared data were collected from the $WISE$ and $IRAS$ catalogs.
To probe the correlation between the two anomalies, we supplemented 15 Li-rich K giants discovered from this sample with
25 known Li-rich K giants from other studies. Dust shell evolutionary models and spectral energy distributions were 
constructed using the code DUSTY to estimate different dust shell properties, such as dust evolutionary time scales, dust
temperatures, and mass-loss rates. }
{Among 2000 K giants, we found about two dozen K giants with detectable far-IR excess, and surprisingly, none
of them are Li-rich. Similarly, the 15 new Li-rich K giants that were identified
from the same sample show no evidence of IR excess.
Of the total 40 Li-rich K giants, only 7 show IR excess.
Important is that K giants with Li enhancement and/or
IR excess begin to appear only at the bump on the RGB.}
{Results show that K giants with IR excess are very rare, similar to K giants with Li enhancement. This may be 
due to the rapid differential evolution of dust shell and Li depletion compared to RGB evolutionary time scales. We also
infer from the results that during the bump evolution,  giants probably undergo some internal changes, which
are perhaps the cause of mass-loss and Li-enhancement events. However, the available
observational results do not ascertain that these properties are correlated.  
That a few  Li-rich giants have IR excess seems to be pure coincidence.}
\end{abstract}

{}


\authorrunning{Bharat}
\titlerunning{Far-infrared properties of K giants}
\maketitle

\section{Introduction}

Standard stellar evolutionary models predict
severe depletion of Li in red giant branch (RGB) stars owing to convection \citep{iben1967a,iben1967b}. 
Contrary to expectations, some RGB stars are found to have enhanced Li in their photosphere. 
Since the discovery of unexpectedly large Li in HD~112127 in 1982 \citep{wallerstein1982},
 a number of Li-rich  K giants ($log \epsilon (Li) \geq 1.5$ dex) have been found \citep{hanni1984,gratton1989}. 
However, systematic surveys among RGB stars have shown that the Li-rich K giants 
are rare, just 1-2$\%$ in the Galaxy  \citep{brown1989,gonzalez2009,bharat2011,lebzelter2012,monaco2011,ruchti2011} and nearby dwarf spheroidal galaxies \citep{kirby2012}.

It is important to know whether the Li-rich K giants contribute to the Galactic Li enrichment.
Enhancement of the Li abundance from its primordial value, A(Li) $\approx$ 2.27 - 2.72 dex \citep{lind2009,cyburt2008}, to
its present level, A(Li) $\approx$ 3.3 dex \citep{matteucci2010},  in the Galaxy is clear evidence
of Li production in the post-big bang nucleosynthesis.  
Some of the known sources of Li enhancement in the universe are  cosmic ray spallation, supernovae, novae, asymptotic giant branch (AGB) stars (\citealt{matteucci2010} and references therein).

Evidence of the RGB contribution to the Galactic Li enhancement
comes in the form of a proposed correlation between Li enhancement and far-infrared excess
in RGB stars \citep{gregoriohetem1993, delareza1996,delareza1997,delareza2012}.
However, these studies were based on a
small number of Li-rich K giants. Most important is that many of the Li-rich K giants discovered in their studies
came from the Pico Dias Survey (or PDS),  which was meant for finding T Tauri stars based on far-IR colors
that prompted more surveys to look for Li-rich
K giants among $IRAS$ sources \citep{delareza1997,castilho1998,fekel1998,
jasniewicz1999}.
Obviously, the number of Li-rich K giants with IR excess among K giants
has increased
\citep{delareza1997,reddy2002, reddy2005}.

Like Li-enhancement, IR excess is also unexpected in RGB stars \citep{zuckerman1995,plets1997}.
Based on dust shell evolutionary models,
\citet{delareza1996} suggested that IR excess
and a large amount of Li  in RGB stars were transient phenomena that would last for a few tens of thousand years.
\citet{palacios2001} provides a mechanism where
rotation-induced mixing  could cause both the Li-enhancement,  as well as mass loss
during their evolution of the RGB luminosity bump.
Another interesting study was done by
\citet{denissenkov2004} and \citet{carlberg2010}, who suggest that Li enrichment
could be due to enhanced extra mixing  triggered by external events, such as the plunging of massive
planets or by a brown dwarf.
According to them, IR excess could be due to a dusty disk orbit around the Li-rich giant (see also, \citealt{jura2003}),
which might have formed either due to the engulfment of planets or the evaporation of planetesimals \citep{plets1997,jasniewicz1999}.
This would mean Li-rich K giants with IR excess could be anywhere on the RGB and not necessarily at the luminosity bump (see \citealt{charbonnel2000,bharat2011}).

In this paper, we have revisited the problem in light of the increased number of Li-rich K giants
that are twice that of the samples involved in the previous studies and the
availability of better astrometry and infrared data
for a large sample of RGB stars. According to our knowledge this is the first uniform study
involving a large sample of 2000 low-mass K giants to search for
evidence of a connection between the two anomalous properties: Li-enhancement and IR excess.

\section{Sample selection and data}
Our sample contains 2000 low-mass (0.8 - 3.0M$_{\sun}$) K giants
in the solar neighborhood ($d \leq$ 200 pc),
with accurate astrometry (parallaxes and proper motions)
taken from the Hipparcos Catalog \citep{vanleeuwen2007}.
This is the same sample as was used by
\cite{bharat2011} for their search for Li-rich K giants.
The sample covers the RGB evolution from its base to well
above the RGB bump, spanning the luminosity range from log(L/L$_{\sun}$) $\approx$ 1.0 to 2.5.
In the survey they found 15 new
Li-rich K giants \citep{bharat2011},
14 of which belong to the thin disk, and one belongs to the thick disk.
In addition to 15 new Li-rich K giants, another  25 were included from the literature following
the criterion suggested 
by \citet{charbonnel2000}. The Li abundance data for the entire Li-rich sample was
taken from
\citet{bharat2011}, \citet{liu2014}, \citet{adamow2014}, and references therein. Of the 25 Li-rich K giants taken from
the literature
22 belong to the thin disk component, one to the thick disk component and one to the halo.
The status is unknown for the remaining star because kinematic velocities are not available.
Li-rich K giants in the sample have solar metallicity within $\pm0.1$dex.

\begin{figure}
 \centering
 \includegraphics[scale=0.7]{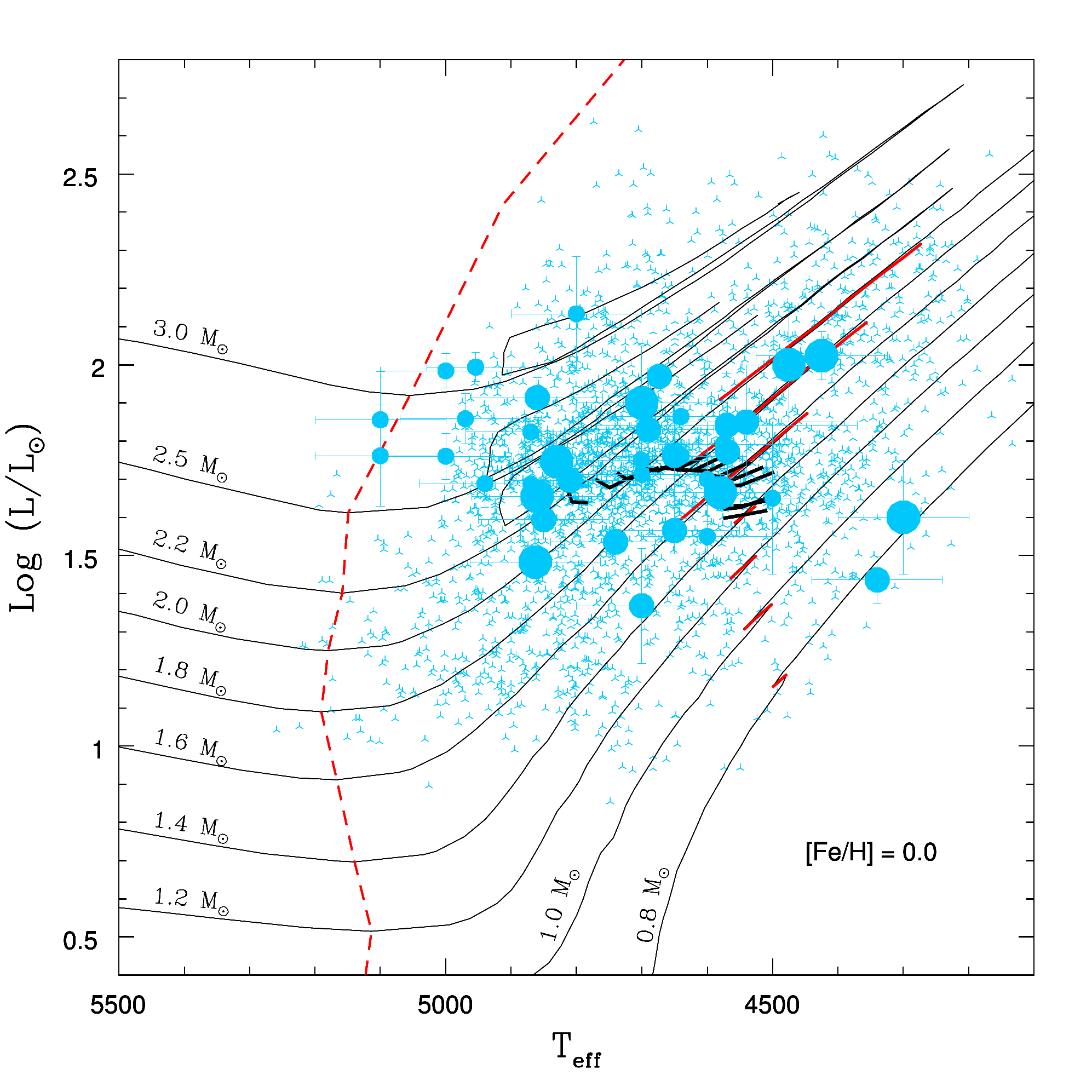}
 \caption{HR-diagram showing survey sample (crosses) and the Li-rich K giants (filled circles)
along with evolutionary tracks for
 low mass stars (0.8 - 3M$_{\sun}$).
The size of the circle indicates amount of Li abundance}
\label{fig1}
\end{figure}

Owing to the unavailability of astrometry and, in some cases, of infrared data, 
Li-rich K giants from the recent surveys
of different populations are not included: Galactic thick disk \citep{monaco2011}, halo \citep{ruchti2011},
bulge \citep{gonzalez2009,lebzelter2012},
and dwarf Spheroidal galaxies \citep{kirby2012}.
We did not include Li-rich giants from other studies \citep{martell2013,monaco2014,silva2014} for the same reason.
The entire survey sample, along with 40 Li-rich K giants, is shown in the Hertzsprung-Russell
(HR) diagram (Figure~\ref{fig1}).
The sample is superposed
on the evolutionary tracks \citep{bertelli2008} of masses from 0.8 to 3.0M$_{\sun}$ with solar
metallicity of [Fe/H] = 0.0.  Also, shown in Figure~\ref{fig1} are the base of the
RGB (dashed red line) and
the region of red clump (black lines) and the luminosity bump (red lines)
on each of the mass tracks.
Li-rich K giants are shown with symbol size indicating amount of Li abundance.

 Infrared data of the entire sample including Li-rich K giants come
from the catalogs of the Wide-field Infrared Survey Explorer ($WISE$) \citep{cutri2012,cutri2013,wright2010} and the Infrared Astronomical Satellite($IRAS$) \citep{helou1988,kleinmann1986,moshir1990}. 
The $IRAS$ catalog contains flux densities
at four different
bands with
central wavelengths of 12$\mu$m, 25$\mu$m, 60$\mu$m, and 100$\mu$m. The
100$\mu$m flux densities are upper limits for most of the sample stars.
For our study, we used
only the first three $IRAS$ bands. 
We divided our sample into three groups: good, moderate, and not good.
The group with measured flux densities in all three bands are considered as good, the group  with measured flux densities
in only two bands are considered as moderate, and the group  with
measured flux densities only in one band is considered as not good.
Of 2000 giants in the sample, we have good $IRAS$ data only for 114 stars and moderate data for 1144 stars.
We have used only good and moderate $IRAS$ data in the analysis.

The $IRAS$ data is supplemented by $WISE$ infrared flux densities. $WISE$ has four IR bands
known as W1, W2, W3, and W4 with central wavelengths
at 3.3$\mu$, 4.6$\mu$, 11.6$\mu$, and 22.1$\mu$, respectively.  
We used flux densities measured at the W2, W3, and W4 bands. Of 2000 sample stars, 1880 have measured flux densities in all three bands, 
and all the sample stars have measured flux densities in W3 and W4. 
All the 40 Li-rich K giants have measured flux densities in W3 and W4 bands, but one star has upper limit 
flux density in the W2 band. We did not include stars with WISE upper-limit data, which is a very small number any way. 
Unfortunately, $AKARI$ far-IR (65$\mu$) data \citep{yamamura2010}
is only available for three stars in our sample. We do not comment any further on this data. 


\section{Far-IR color-color diagram}
Far-IR (FIR) flux excess indicates the presence of
circumstellar dust, which can be traced using the FIR color-color diagram (FIR-CCDm).
The FIR colors between the two bands with central wavelengths
$\lambda_{1}$ and $\lambda_{2}$ were calculated using the following equation:
\begin{equation}
 [{\lambda_{1}} - {\lambda_{2}}] = log ({\lambda_{2}}{f_{\lambda_{1}}}) - log ({\lambda_{1}}{f_{\lambda_{2}}})
\label{eq1}
\end{equation}

\noindent where
${f_{\lambda_{1}}}$ and ${f_{\lambda_{2}}}$ are the flux densities.

\subsection{Survey sample}

$IRAS$ colors of the sample stars of  good and moderate groups are shown in Figure~2a, along with Li-rich K giants. 
Following the study of $IRAS$ data of stars with dust envelopes by \citet{vanderveen1988},  we partitioned the CCDm into different boxes and labeled them accordingly.
 The Vanderveen diagram is found to be useful tool for interpreting the evolution of dust around the star. For example,  Region I
represents stars
with no dust, and Regions II, IIIa, IIIb, IV, VIb, and VIa represent stars
with dust going progressively from warm to cool.
Photospheric far-IR colors of RGB
stars with no IR excess are represented by Region $RGB$ (see Figure~2a).
As shown in Figure~2a,
most of the stars are in Region $RGB$ and VIa, and a very
few are in other regions. Of the stars with good data, the majority
are in Region $RGB$ with no excess of dust. A few are in Regions VIa and VIb.
However, the vast majority of
stars with moderate data are in Region VIa. This could be due to
60$\mu$m flux density, which is an upper limit. We assume that all of these stars
belong to Region $RGB$ with no excess of dust.
We note that only a small fraction of stars with good data
show detectable far-IR excess, which confirms the earlier estimation
of far-IR excess among the giants of luminosity
class III (see \citealt{zuckerman1995} \& \citealt{plets1997}).

A similar diagram is shown in Figure~2b for two $WISE$ measurements 
at 11.6$\mu$ (W3) and 22.1$\mu$ (W4) and $IRAS$ 60$\mu$ flux density. We have
$WISE$ measured flux densities of [22-11] for 1880 sample stars, including all the 15 Li-rich 
K giants from the sample against $IRAS$ measured flux densities of [25-12] for 1144 sample stars.
$WISE$ colors [22-11] show much less scatter 
and show excess for very few stars. Also, none
of the 15 Li-rich K giants show excess in [22-11] and fall in the same region
as normal K giants, confirming the $IRAS$ data based on a relatively small number 
of measured flux densities.

\subsection {Li-rich K giants}
We have  a total of 40 Li-rich K giants,  15 of which come from our survey \citep{bharat2011}, and
the rest are from the literature. All of them are shown in Figure~3a of $IRAS$ CCD.  
Arcturus, a typical K giant with
very low Li (A(Li)$\sim$-0.6) and no infrared excess, is shown
as a reference. The $IRAS$ data and colours of 40 Li-rich K giants are given in
Table~\ref{tbl1}.
Of 40 Li-rich K giants, 7 have ``good data'', 18 have ``moderate data'', 
and another 15 have ``not good data''. Unfortunately, none of the 15 Li-rich K giants
found from our survey of 2000 K giants belong
to the good
data group. However, six of them have measured flux densities in 12$\mu$ and 25 $\mu$ and fall in the region VIa.
From Figure~3a, we note only 7 Li-rich K giants showing IR excess in
both 25 $\mu$ and 60 $\mu$ bands, suggesting clear evidence of a dust envelope 
and of the recent  mass loss. 
For 18 stars with moderate data, flux densities of 12 $\mu$ and 25 $\mu$ bands are 
normal and show no evidence of 
excess flux. However, [60-25] color indicates cool dust (region VIa), 
but this may be entirely due to the upper limit in 60 $\mu$ flux.  These candidates most likely 
show black-body [25-12] color of their photospheres, and at 60$\mu$ the 
photospheric flux densities are too faint to be detected. The third group (black symbols)
have only upper limit flux densities both in 25 $\mu$ and 60 $\mu$ and would be difficult to say
whether they have or do not have IR excess. 

However,  
$WISE$ measured flux densities at 22.1$\mu$ \& 11.6$\mu$ are available 
for all 40 Li-rich K giants. The good quality of $IRAS$ 60 $\mu$ flux densities are 
available for only seven Li-rich K giants, and the rest are with upper limits. In  
Figure~3b, we showed
all the 40 Li-rich K giants in the plot of [22-11] and [60-22] color. It is clear from the
figure that stars that have measured flux densities in all three bands are the same 
seven Li-rich K giants that show IR excess indicating some kind of warm dust. The
position of black symbols in Figure~3a moved to the left in Figure~3b, showing no IR excess 
in [22-11] color. Since 60$\mu$ flux densities are upper limits, we assume that all the Li-rich K giants
represented in blue are similar to normal K giants with expected photospheric fluxes. Also, in Figure~4 we showed $WISE$ colors for the
entire sample (1880) and 39 Li-rich K giants excluding 
stars with upper limits. This too confirms
the above findings.
\section {Modeling of circumstellar\ envelope: DUSTY}

 We used a 1-D radiative transfer code,  DUSTY \citep{ivezic1999}, to model
the observed spectral energy distribution (SED) and to reconstruct
the $IRAS$ color-color diagram with dust evolutionary models.
DUSTY solves radiative transport through a dusty region in the
circumstellar environment. The solution is obtained through
an integral equation for the spectral  energy
density. DUSTY can handle
the geometry of the regions that are plane
parallel or spherical density distribution heated by a central source.  It has built-in
optical constants for most of the astronomical dust grains. 
\subsection{Spectral energy distribution (SED)}
Modeling of SED,  constructed from  far-UV to the sub-mm data of
dust shells around cool giants, is one of the important diagnostic tools for understanding
the nature of the dust.
To construct the SED, we took optical (BVRI) \citep{monet2003}, 
near-infrared (JHK$_{s}$) magnitudes \citep{skrutskie2006}, and far-infrared ($IRAS$, $WISE$) 
flux densities \citep{helou1988} from
the respective catalogs and
SIMBAD.
Mid-infrared fluxes observed by MSX \citep{egan2003} were taken wherever applicable.
Magnitudes are converted
to fluxes (see \citealt{bessel1998,cohen2003}).  The model SED is fitted to 
the observed SED by scaling
the K band model flux to the observed value. In the modeling of dust envelope we used only
measured fluxes, not the fluxes with upper limits.

DUSTY computes SEDs for a given set of input parameters: dust temperature at the 
inner shell (T$_{d}$), optical depth ($\tau$) at 0.55$\mu$, the relative 
shell thickness (R$_{out}$-R$_{in}$)/R$_{in}$,
and the input SED of the central star. The central star was taken as a point source with a spherically distributed dust. To compute the SED for a central star, we have
used appropriate model atmospheres from a Kurucz grid of model atmospheres \citep{kurucz1994}.
Atmospheric parameters (T$_{eff}$, log $g$, and [Fe/H]) were taken from \citet{bharat2011}.
The dust shell was assumed to have astronomical silicate grains because most of the RGB stars are
oxygen-rich (C/O $\leq$ 1.0).
The dust temperature at the inner shell, optical depth, and relative shell thickness were found by an
iterative method until the theoretical
SED matched the observed
SED well. We found that the excess emission was reproduced well with the warm silicate dust
grains \citep{ossenkopf1992}.
Derived  dust temperatures (T$_{d}$) along with the atmospheric parameters for Li-rich giants with good data are given in
Table~\ref{tbl2}.

\subsection{Mass-loss rates and envelope kinematic age}
Dust parameters derived from modeling SEDs are used to derive mass-loss rates and kinematic age.
Assuming a constant velocity (V$_{s}$) of 2 km s$^{-1}$ for the wind, the rate of mass loss and
kinematic age ($t$) of the dust shell are estimated following
 \citet{delareza1996}:
\begin{equation}
\dot{M} =  {7.9{\times}10^{-28}}{\tau_{v}}{R_{in}^{'}}{V_s}; \\    
t = \frac{R_{in}^{'}}{V_s}
\end{equation}
where $\tau_{v}$ is the optical depth at visible band, R$_{in}^{'}$ is inner dust shell radius.
The derived values for the seven Li-rich K giants with good data are given
in Table~\ref{tbl2}.

Furthermore, we predicted mass loss rates of RGB stars at different phases in their evolution.
In Table~\ref{tbl3}, we have provided expected mass-loss
rates.
The mass-loss rates are computed using
the modified version of the Reimer's law \citep{schroeder2005} ($\dot{M}_{\rm R}$):

\begin{equation}
\dot{M_{\rm R}} = {\frac{{\eta}{L_\star}{R_\star}}{M_\star}} (\frac{T_{eff}}{4000}) (1+{\frac{g_{\sun}}{4300g_\star}}) 
\end{equation}

where $\eta$ = 8$\pm$1$\times$10$^{-14}$ M$_{\sun}$ yr$^{-1}$; L$_\star$, R$_\star$, M$_\star$, $g_\star$ are luminosity, radius, mass, and
surface gravity of a given star, respectively. For solar gravity, log $g_{\sun}$ = 4.44 has been adopted.
Radii for the stars are calculated from their luminosities and temperatures. Stellar gravities
(log g$_{\star}$) have been derived
using the following standard equation:

\begin{equation}
log{g} = log{\frac{L}{L_{\sun}}} + log{\frac{M}{M_{\sun}}} + 4log{T_{eff}} - 10.61
.\end{equation}
The expected mass-loss rates of normal giants are in the range
of 10$^{-12}$ to 10$^{-10}$ M$_{\sun}$ yr$^{-1}$, which are one to two orders of magnitude
lower than the observed values of Li-rich K giants with far-IR excess (Table~\ref{tbl2} \& Table~\ref{tbl3}).

\subsection{Circumstellar dust shell evolutionary model}
In Figure~\ref{fig5}, DUSTY-computed far-IR colors are shown for a set of stellar parameters
covering RGB evolution. For computation, we assumed that the dust shell is spherically symmetric, thin, and detached
from the central star and that it is a uniform mixture of dust and gas.
 We also assumed that the newly formed dust
shell is a result of uniform mass loss, hot with T$_{d}$ = 1500 K  (sublimation temperature of silicates) and  cooling down to $T_{d}$ = 15 K ( dust temperature of diffuse interstellar medium) before it dissipates into the interstellar medium. In computing the dust shell evolution, we considered
four different mass-loss rates: 1$\times$10$^{-10}$, 1$\times$10$^{-9}$, 5$\times$10$^{-9}$,
and 3$\times$10$^{-8}$.

\section {Discussion}
From the survey, we found only 114 K giants with measured flux densities in all three $IRAS$ (12$\mu$, 25$\mu$, \& 60$\mu$) bands. 
Of these only
23 show evidence of far-IR excess due to cold dust (regions V1a, VIb), and the rest have 
no excess of dust (region $RGB$) (see Figure~2a).
Interestingly, none of these K giants with far-IR excess are Li-rich. Similarly, only 15 out of the 2000 K giants are Li-rich (see \citealt{bharat2011}),
and none of them show evidence
of far-IR excess. For many stars in the sample, only $IRAS$ upper limits are available for both 12$\mu$ and 25$\mu$ . 
To improve the situation we supplemented $IRAS$ data with $WISE$ data that has measured flux densities 
for almost all the stars in 11.6$\mu$ and 22.1$\mu$.   
The combined $IRAS$ and $WISE$ data results suggest that
the far-IR excess among  K giants is uncommon, which confirms earlier results by
\citet{zuckerman1995} and \citet{plets1997}.  This may be due to the fast depletion of enhanced Li 
and dust dissipation in an uncorrelated way. Important is that
the unbiased survey results do not 
confirm earlier suggestions \citep{delareza1996,delareza1997}
of a correlation between the two anomalous properties of K giants: Li-enhancement and far-IR excess.

To gain more insight, we included 25 Li-rich K giants from the literature. These are shown in Figure~3a along with
15 (black \& blue triangles) Li-rich K giants from  the present survey. Of 40, only
7 have measured flux densities in all the three $IRAS$ (12$\mu$, 25$\mu$, and 60$\mu$) bands, and show evidence of warm dust. 
Among seven super Li-rich K giants  (log $\epsilon$ Li $\geq$ 3.3) in the sample only three
show IR excess and all the three have been selected based on IR colours. Important is that four 
other super Li-rich K giants show no evidence of IR excess, certainly no warm dust around giants with such high Li
(Figure~3b). This observational result contradicts the hypothesis \citep{delareza1996,delareza1997} that the evolution of Li 
abundance and IR excess
are correlated wherein
the larger  the Li abundance in K giants, the hotter the circumstellar dust, the less the Li abundance, and the cooler the circumstellar dust.
The large Li abundance in K giants implies recent enhancement of Li 
accompanied by a mass-loss event (hence the warm dust), suggesting that not much time has elapsed since the events that
caused Li enhancement and dust shell ejection. 

Results from dust models, shown in Figure~\ref{fig5}
and Table~\ref{tbl4}, suggest that the dust evolutionary time scales are quite small at a few thousand to a few hundred
thousand years for typical K giants with the least mass loss rates (1 $\times 10^{-10}$) to heavy
mass loss rates (5 $\times 10^{-8}$), respectively.
Also, Li-depletion timescales, according to Li Flash model \citep{palacios2001}, are on the order of 10$^{4}$ ($\approx$ 20000) yrs.
These time scales are much smaller than stellar evolutionary time scales \citep{bertelli2008}
and, important, smaller than the evolutionary timescales of different phases along the RGB (see Table~\ref{tbl5}).
K giants spend a few million years ($t_{bump}$) at the bump
and take about 10-100 million years to evolve from the bump to the clump via the tip of the RGB.
The above arguments suggest that the ejected dust shell due to mass loss and the enriched Li may not even
survive the bump
evolution. 
It is still an open question whether it is
just a coincidence that seven Li-rich K  giants have been found to have far-IR excess (Figure~3a,3b) or that these two properties
are somehow related but not observed
owing to their fast but different evolutionary scales. To understand the later question, we plotted
the entire sample along with the 40 Li-rich K giants in the HR diagram ( Figure~\ref{fig6}).
Results shown in Figure~\ref{fig6} complicate the situation further.
From close examination of the displayed data, one could 
make a few interesting observations: a) there seem to be two groups of
Li-rich K giants:  one at the
luminosity bump and another at the clump, b) the Li-rich K giants
that overlap with the clump region (just below the bump and to the left)
do not show far-IR excess, c) some of the Li-rich K giants in the bump region
show far-IR excess due to warm dust, d) we note that the K giants with
far-IR excess and/or Li-enhancement
occur in the narrow luminosity range (log(L/L$_{\sun}$) = 1.4 - 2.2).

The absence of
Li-rich K giants below the bump suggests that
Li production begins during the bump evolution. Interestingly,
the K giants with far-IR excess also begin to appear
at or within the bump region. 
In fact, all seven Li-rich K giants with far-IR excess
are in the bump region (Figure~\ref{fig6}).
However, within the same region, there are also Li-rich giants without far-IR excess  and far-IR excess K giants
without Li enhancement. Thus, one ought to be cautious
about associating Li-enhancement in K giants with the mass-loss events or vice versa.
 Also, assuming the possibility that planet engulfment enriches host-star Li abundance, hence IR excess ( See \citealt{denissenkov2004,carlberg2010,adamow2014}), and that the engulfment may happen anywhere along the RGB, one would expect Li enhancement across the RGB but not necessarily at the narrow bump luminosity. Absence of K giants with Li enhancement below and above the bump region casts shadow on the external origin scenario.

The presence of Li-rich and also super Li-rich K giants without far-IR excess
(blue triangles in Figure~3a, box~3) at the clump (see green shade in Figure~\ref{fig6})
requires an alternative site
for Li enhancement, other than the bump.  It is very unlikely that the Li produced at the bump remains
at its peak value.  As giants evolve
to the clump via the tip of the RGB they experience deep convection.
The extent of mixing is evidenced by their very low values of $^{12}$C/$^{13}$C ratios
(see \citealt{bharat2009}; \citealt{bharat2011}). Thus,
in \citet{bharat2011} study core-He flash
at the tip of the RGB has been suggested as the most likely alternative
event that could cause Li
enhancement that might or
might not have been associated with the mass loss. Since the
evolution from the tip of the RGB
to the clump is rapid, Li produced at the tip is expected to survive. However,
in a recent study by \citet{denissenkov2012},
an alternative scenario was proposed in which, as a result of extra-mixing, giants originated
in the bump  could
make extended zigzags that
may reach luminosities below the bump luminosity explaining a group
of Li-rich K giants that appear
to overlap with the red clump in the HR diagram (Figure~\ref{fig6}).

\section{Conclusion}
We performed  a search for a correlation between Li-enhancement and far-IR excess
among  a uniform sample of 2000 K giants. Infrared data for the sample is taken from the $WISE$ and $IRAS$ catalogs.  
Results suggest that IR excess similar to Li enhancement among
the K giants is uncommon.
Results from the survey
show no direct evidence of correlation between the two anomalous properties of K giants. None of the
15 Li-rich K giants show IR excess, and none of the K giants that have measurable IR excess show Li enhancement.

Furthermore, Li-rich K giants from the survey are supplemented by 25 known Li-rich K giants from different studies.
Absence of Li-rich and/or
IR excess K giants below the bump stresses the fact that K giants undergo some key internal changes
during the short span of bump evolution leading
to mass loss and Li enhancement. Though the evidence for
the two events  occurring during the bump are clear from our study, what is not clear is
whether one phenomenon triggers another,
leading to a correlation, or  whether the two events are independent. Results from our observations and from
the compilation of Li-rich K giants favor the later suggestion.
The presence of both IR excess and Li enhancement in a few K giants seems to be a coincidence.

To get a better perspective on the problem, it is important to search for Li enhancement among
red clump stars and RGB bump stars separately.  Also, several studies \citep{gonzalez2009,monaco2011,lebzelter2012,ruchti2011,martell2013,adamow2014} report that some giants with large
Li enhancement are occupied between the bump and tip of the RGB. It would be a worthwhile exercise to differentiate
RGB stars from the early asymptotic giant branch (AGB) stars. 
It is well known that many AGB stars produce Li
through the Cameron-Fowler mechanism \citep{cameron1971} in their interiors and gets dredged up to the surface 
(see \citealt{sackmann1992}).

\begin{acknowledgements}
We sincerely thank the referee who made constructive suggestions of using $WISE$ data, which made our interpretations more robust.
This study is sponsored by the Chinese Academy of Sciences Visiting Fellowship for Researchers from Developing Countries, Grant No. 2013FFJB0008, and partially supported by the National Natural Science Foundation of China under grants Nos. 11450110404, 11390371, and 11233004. Y.B.K is thankful to T. Medupe for his kind support through NRF grant at NWU.
This research made use of the Simbad database and the NASA ADS service.
\end{acknowledgements}

\begin{table}
\caption{ $IRAS$ flux densities and Li abundances of Li-rich giants.
\label{tbl1}}
\flushleft
\begin{tabular}{lccccccccrrr}
\hline
\hline
Star & A(Li) & F12 & Quality & F25 & Quality & F60 & Quality & [25-12] & [60-25] & Reference \\
&  & (Jy) & & (Jy) & & (Jy) & &&  \\
\hline
        HD183492 &  2.07  &   2.377 & 2 & 2.512  & 1 &  0.8849 & 1 &    -0.294  & -0.833 & 3 \\
        HD107484 &  2.14  &   0.410 &    3 &       0.156 &    1 &     0.298 &    1 &         -0.738  &   -0.099 & 7  \\
        HD108471 &  2.10  &   0.869 &    3 &       0.361 &    1 &       0.4 &    1 &         -0.700 &   -0.335  &  3 \\
        HD118319 &  2.02  &    1.04 &    3 &       0.287 &    1 &       0.4 &    1 &         -0.877 &   -0.236  &  7 \\
        HD120602 &  2.07  &   1.129 &    3 &       0.278 &    1 &       0.4 &    1 &         -0.927 &   -0.222  & 3 \\
         HD8676  &  3.55  & 0.319 &    3 &       0.188 &    1 &     0.140 &    1 &         -0.548 &   -0.508  & 7 \\ 
        HD12203  &  2.08  & 0.793 &    3 &       0.409 &    1 &       0.4 &    1 &          -0.606 &   -0.389 & 7  \\
        HD133086 &  2.14  & 0.706 &    3 &        0.25 &    1 &       0.4 &    1 &         -0.769 &   -0.176  & 7 \\
         HD6665   & 2.93    &    0.292  &   3 &       0.192 &    1 &      0.163 &   1 &         -0.501 &   -0.451 & 13 \\
         HD217352 & 2.65   &   0.701 &    3 &       0.308 &    1 &       0.4 &    1 &         -0.675 &   -0.266  & 13 \\
         HD63798 &  2.00   &   0.865 &    3 &       0.347 &    1 &       0.4 &    1 &         -0.715 &   -0.318  & 10 \\
         HD88476 &  2.21  &   0.638 &    3 &        0.29 &    1 &       0.4 &    1 &          -0.661 &   -0.240 & 7  \\
        HD150902 & 2.65   &    0.298 &    3 &        0.25 &    1 &       0.4 &    1 &         -0.396 &   -0.176  & 7 \\
         HD37719 &  2.71  &   0.376 &    3 &        0.25 &    1 &       0.4 &    1 &         -0.496 &   -0.176  & 7 \\
         HD40168 &  1.70  &   0.759 &    3 &        0.25 &    1 &       0.4 &    1 &         -0.801 &   -0.176  & 7 \\
        HD203136 & 2.34   &   0.444 &    3 &       0.305 &    1 &     0.584 &    1 &         -0.482 &   -0.098  & 13 \\
 TYC3105-00152-1 & 2.86   &  0.0602 &    3 &     0.0510  &  1  &  0.0806 &   1 &          -0.391 &  -0.181 &  1 \\   
        HD112127 &  2.95   &  0.979 &    3 &      0.290 &    2 &       0.4 &    1 &          -0.846 &    -0.241 & 14  \\
         HD40827 &  2.05  &    1.51 &    3 &       0.368 &    2 &       0.4 &    1 &         -0.931 &   -0.343  & 3 \\
        HD170527 &  3.12  &   0.873 &    3 &     0.264   &    2 &       0.4 &    1 &         -0.837 &   -0.200   & 7\\
         HD51367 &  2.60 &   0.705 &    3 &     0.246  &    2 &       0.4 &    1 &          -0.774 &    -0.170  & 7 \\
         HD10437 &  3.48 &     1.12 &    3 &     0.304 &    2 &       0.4 &    1 &          -0.884 &     -0.261  & 7 \\
        HD212430 & 1.83  &    1.81 &    3 &      0.385 &   2 &      0.4 &   1 &            -0.991 &   -0.364 &  9 \\
        HD102845 & 1.98  &   1.08  &  3   &    0.207 &   2 &      0.4 &   1 &              -1.036 &  -0.094 &  9 \\ 
        HD145457 &   2.49   &  0.988 &    3 &     0.316 &    3 &       0.4 &    1 &          -0.813 &     -0.278  & 7 \\
        HD167304 &   2.85   &   1.28 &    3 &       0.281 &    3 &       0.4 &    1 &       -0.977 &   -0.226   & 7 \\
        HD214995 &   2.95    &   1.84 &    3 &       0.438 &    3 &       0.4 &    1 &       -0.942 &   -0.419   & 8 \\
         HD90633 &  2.18     &    1.41 &    3 &       0.346 &    3 &       0.4 &    1 &       -0.928 &   -0.317   & 10 \\
        HD148293 &  2.16     &    3.52 &    3 &     0.78744 &    3 &       0.4 &    1 &       -0.969 &   -0.674   & 3 \\
        HD116292 &  1.50 &   2.51 &    3 &       0.661 &    3 &       0.4 &    1 &       -0.898 &   -0.598   & 3 \\
          HD9746 & 3.44  &       2.74 &    3 &       0.712 &    3 &       0.4 &    1 &       -0.904 &   -0.630   &  3 \\
         HD77361 &  3.80 &      1.58 &    3 &       0.427 &    3 &       0.4 &    1 &       -0.886 &   -0.408   & 7 \\
        HD194937 & 3.18  &      1.51 &    3 &       0.440 &    3 &       0.4 &    1 &       -0.855 &   -0.420   & 8 \\
  $IRAS$17596-3952 &  2.30  &   0.482 &    3 &        1.11 &    2 &     0.637 &    2 &          0.044 &   -0.621  & 12 \\
         HD19745 &  3.40  &   0.332 &    3 &       0.777 &    3 &     0.612 &    3 &           0.050 &   -0.483 & 12  \\
  $IRAS$13539-4153 &  3.90 &   0.491 &    3 &       0.821 &    3 &     0.644 &    3 &          -0.095 &   -0.485 & 12  \\
  $IRAS$13313-5838 & 3.13   &     1.88 &    3 &        6.04 &    3 &       3.3 &    2 &          0.188 &   -0.642  & 4 \\
          PDS100 & 2.40   &     4.74 &    3 &        6.53 &    3 &      3.38 &    3 &         -0.179 &   -0.666  & 11 \\
        HD219025 & 2.93   &     17.1 &    3 &        10.3 &    3 &      3.86 &    3 &         -0.538 &   -0.806  & 5,6 \\
        HD233517 &  3.95  &   0.502 &    3 &         3.6 &    3 &       7.6 &    3 &          0.536 &   -0.055  & 2 \\
\hline
\end{tabular}
\vskip 0.1cm
\tablefoot{1 = Upper limits; 2 = Moderate; 3 = High} \\
\tablebib{(1) \citet{adamow2014}; (2) \citet{balachandran2000}; (3) \citet{brown1989};
 (4) \citet{drake2002}; (5) \citet{fekel1998}; (6) \citet{jasniewicz1999}; (7) \citet{bharat2011}; (8) \citet{luck2007}; (9) \citet{liu2014}; (10) \citet{mishenina2006}; 
(11) \citet{reddy2002};  (12) \citet{reddy2005}; (13) \citet{strassmeier2000}; (14) \citet{wallerstein1982}  
}
\end{table}

\begin{center}
\begin{table}[ht]
\small
\caption{ Photospheric and dust parameters of Li-rich K giants with IR excess.
\label{tbl2}}
\centering
\begin{tabular}{lrrccrcrrr}
\hline\hline
Star & [Fe/H] & T$_{eff}$  & log$g$ & Log(L/L$_{\sun}$) & R/R$_{\sun}$ & {\.{M}} & T$_{dust}$ & t & \\
& & (K) & & & & (M$_{\sun}$yr$^{-1}$) & (K) & (yr) & \\
\hline
  HD 233517 &   $-$0.37     &      4475 &  2.25 &    2.00 &   16.6655 & 1.88577E-07 &        75 &   1262.99 & \\ 
  HD 219025 &   $-$0.10     &      4570 &  2.30 &    1.88 &   13.9179 & 6.02185E-09 &       275 &   40.3312 & \\
  HD 19745 &    $-$0.05     &      4700 &  2.25 &    1.90 &   13.4651 & 3.50690E-09 &       170 &   117.437 & \\ 
   $IRAS$ 13539-4153  &  $-$0.13     &  2.25 &    4300 &      1.60 &   11.3885 & 3.27951E-08 &       230 &   47.0666 & \\ 
    $IRAS$ 17596-3952  &  0.10     &  2.50 &    4600 &      1.70 &   11.1658 & 2.37905E-08 &       190 &   85.3588 & \\ 
    $IRAS$ 13313-5838  &  $-$0.09      &  2.20 &    4540 &      1.85 &   13.6236 & 1.29749E-07 &       260 &    46.553 & \\ 
    PDS 100 &   0.14      &      4500 &  2.50 &    1.70 &   11.6676 & 1.48663E-08 &       250 &     29.87 & \\
 \hline
\end{tabular}
\end{table}
\end{center}

\begin{table}
\caption{Expected Reimers' law mass-loss rates at different locations
on RGB 
\label{tbl3}}
\centering
\begin{tabular}{ccccccrrrrr}
\hline
\hline
Bin & T$_{eff}$ & log(L/L$_{\sun}$) & M$_{\star}$ & R$_{\star}$ & log $g$ & {\.{M}$_{R}$} & \\
 & (K) & & (M$_{\sun}$) & (R$_{\sun}$) & & (M$_{\sun}$yr$^{-1}$) & \\
\hline
Prebump &  4650  &  0.9  &  1.0 &    4.3501  &    3.163    &    4.83757E-12        \\
 & 4650  &  1.5  &  1.6 &   8.67958  &  2.76712    &     2.40163E-11       \\
 & 5200  &  1.0  &  1.9 &     3.903  &  3.53596    &    2.87589E-12        \\
 & 5200  &  1.5  &  2.5 &   6.94062  &  3.15514    &   1.2291E-11          \\
 & 4800  &  1.3  &  1.8 &   6.47027  &  3.07343    &      1.0041E-11       \\
\hline
Bump &   4500 &   1.0  &  0.8 &    5.2117  &  2.90913    &     9.12048E-12     \\
 & 4500 &   1.6  &  1.2 &   10.3987  &  2.48522    &     4.82977E-11      \\
 & 4650 &   1.0  &  1.0 &   4.88089  &    3.063    &     6.83324E-12      \\
 & 4650 &   2.0  &  2.5 &   15.4347  &  2.46094    &     8.64344E-11     \\
 & 4200 &   2.0  &  0.8 &   18.9193  &  1.78928    &     3.31089E-10      \\
 & 4600 &  1.75  &  1.8 &   11.8274  &  2.54949    &     5.17303E-11      \\
\hline
Clump &   4650  &  1.5  &  1.4  &  8.67958  &  2.70913     &  2.74472E-11     \\
 & 4650  &  2.0  &  2.2  &  15.4347  &  2.40543     &   9.8221E-11     \\
 & 5000  &  1.5  &  2.5  &  7.50697  &  3.08701     &  1.32939E-11     \\
 & 5000  &  2.0  &  3.5  &  13.3495  &  2.73314     &   5.3398E-11     \\
 & 4750  & 1.75  &  2.3  &  11.0922  &  2.71169     &   3.7968E-11     \\
\hline
Postbump &   4200  &  2.0  &  1.0  &  18.9193  &  1.88619   &  2.64871E-10    \\
 & 4200  &  2.5  &  2.1  &  33.6439  &  1.70841    &  7.09275E-10     \\
 & 5000  &  2.0  &  3.0  &  13.3495  &  2.66619    &  6.22976E-11     \\
 & 4800  &  2.4  &  4.0  &  22.9574  &  2.32022    &  2.01832E-10    \\
 & 4300  &  2.1  &  1.8  &   20.252  &  2.08234    &    1.983E-10     \\
 & 4500  &  2.2  &  2.5  &  20.7482  &  2.20398    &  1.84148E-10     \\
\hline
\end{tabular}
\end{table}

\begin{center}
\begin{table}
\caption{Dust parameters and evolutionary timescales for different photospheric parameters of RGB stars
\label{tbl4}}
\centering
\begin{tabular}{crrcrr}
\hline
\hline
\.{M} & T$_{\star}$ &  R$_{\star}$ & T$_{dust}$ &  Age & \\
(M$_{\sun}$yr$^{-1}$) &  (K) & (R$_{\sun}$) & (K) &  (yr) &\\
\hline
1$\times$10$^{-10}$ & 4000 & 10 & 25 & 16548 &  \\
& & 20 & 28 & 23641 & \\
& 5000 & 10 & 27 & 37802 & \\
& & 20 & 31 & 50164 & \\
1$\times$10$^{-9}$ & 4000 & 10 & 17  &  52163   & \\
& & 20 & 19  & 74736 & \\
& 5000 & 10 & 19   & 107675   & \\
& & 20 & 20.5  & 171636 & \\
5$\times$10$^{-9}$ & 4000 & 10 &  15   &  75957   & \\
& & 20 &  16  & 125070    &\\
& 5000 & 10 & 15.5  & 198868   & \\
& & 20 & 17  & 300273   & \\
3$\times$10$^{-8}$ & 4000 & 10 & 13   &  116681  & \\
& & 20 & 13.5 & 208958   & \\
& 5000 & 10 & 13.5  & 299198   & \\
& & 20 & 14   & 537481    & \\
\hline
\end{tabular}
\end{table}
\end{center}

\begin{center}
\begin{table}
\caption{Evolutionary timescales from the RGB bump to the red clump  
\label{tbl5}}
\centering
\begin{tabular}{lrrrrrr}
\hline
\hline
M$_{\star}$ & t$_{bump}$ & t$_{bump-tip}$ &  t$_{clump}$ \\
(M$_{\sun}$) & (yr) & (yr) & (yr) \\
\hline
       1.0 & 1.52E+07 &  1.27E+08  &     9.49E+07 \\
       1.2 & 1.11E+07 &  9.87E+07 &      7.98E+07 \\
       1.4 & 8.12E+06 &  7.50E+07  &     7.61E+07 \\
       1.6 & 6.85E+06 &  5.68E+07  &     8.07E+07 \\
       1.8 & 4.07E+06 &  3.54E+07  &     9.17E+07 \\
       2.0 & 2.45E+06 &  1.95E+07  &     1.03E+08 \\
\hline
\end{tabular}
\end{table}
\end{center}

\newpage

\begin{figure}[ht]
\centering
\includegraphics[scale=0.5]{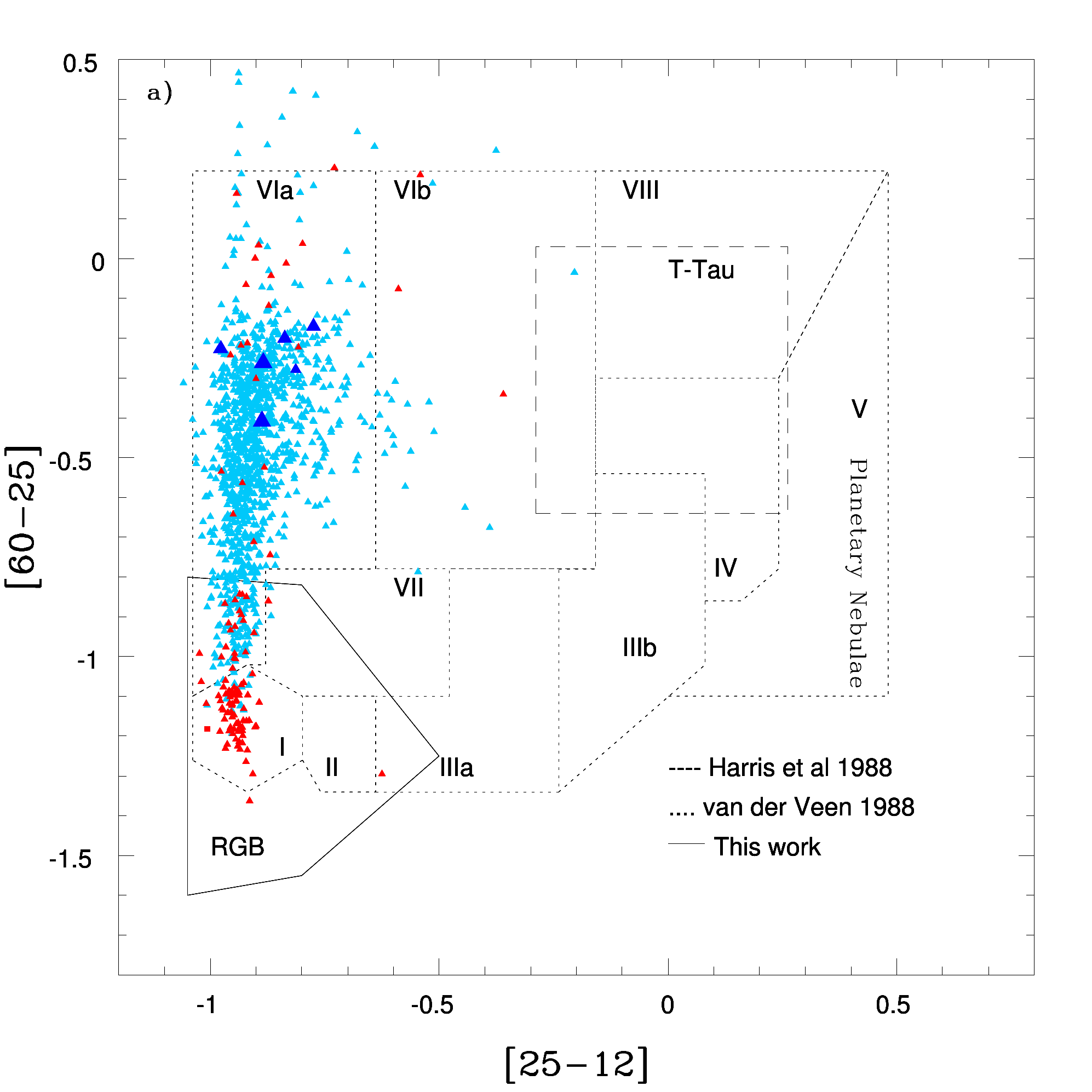}
\includegraphics[scale=0.5]{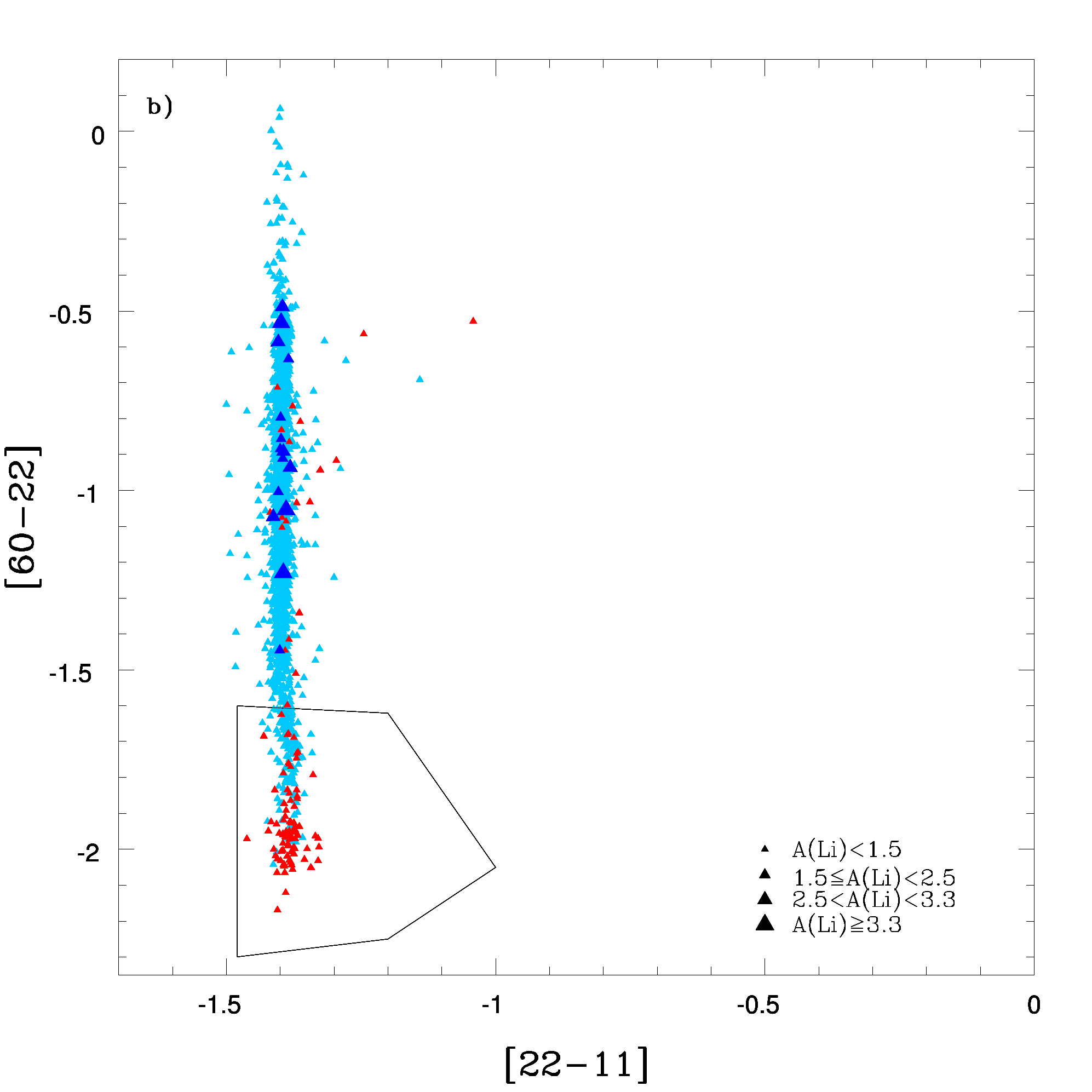}
\caption{Sample stars (triangles) are shown in far infrared color-color diagram.
Red and blue colors indicate good and moderate data, respectively.
Boxes with dotted lines in Figure~2a show different
phases of dust shell evolution based on the $IRAS$ colors. 
The box with a dashed line represents a region of young T-Tauri stars \citep{harris1988}.
The box with a solid line in both Figs.~2a \& 2b represents the region of RGB stars.
Also, shown is  Arcturus, a typical K giant (red square). In Figure~2b, the sample is plotted using $WISE$ 11.6$\mu$ and 22.1$\mu$ and 
$IRAS$ 60$\mu$ flux densities.}
\label{fig2}
\end{figure}

\begin{figure}[ht]
\centering
\includegraphics[scale=0.5]{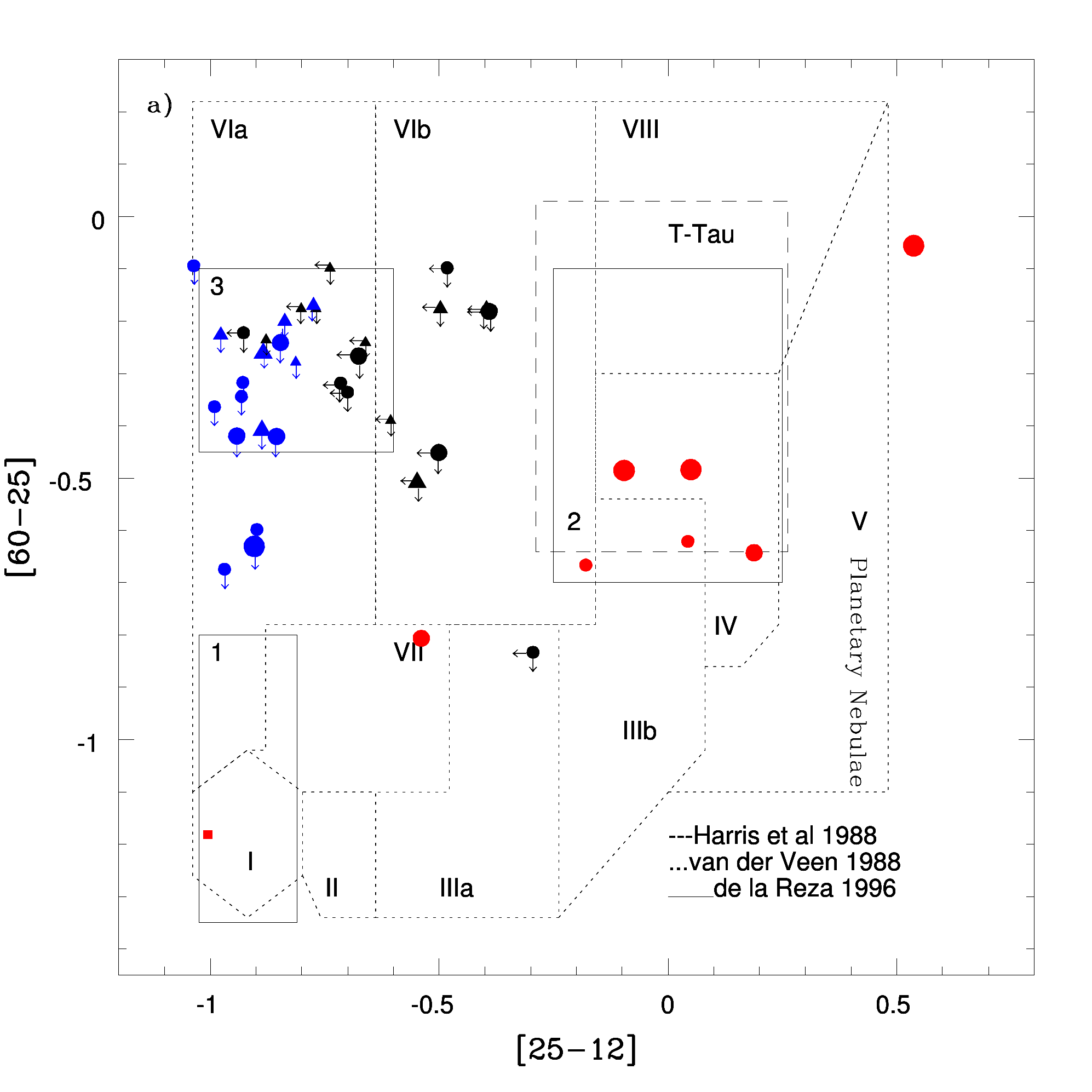}
\includegraphics[scale=0.5]{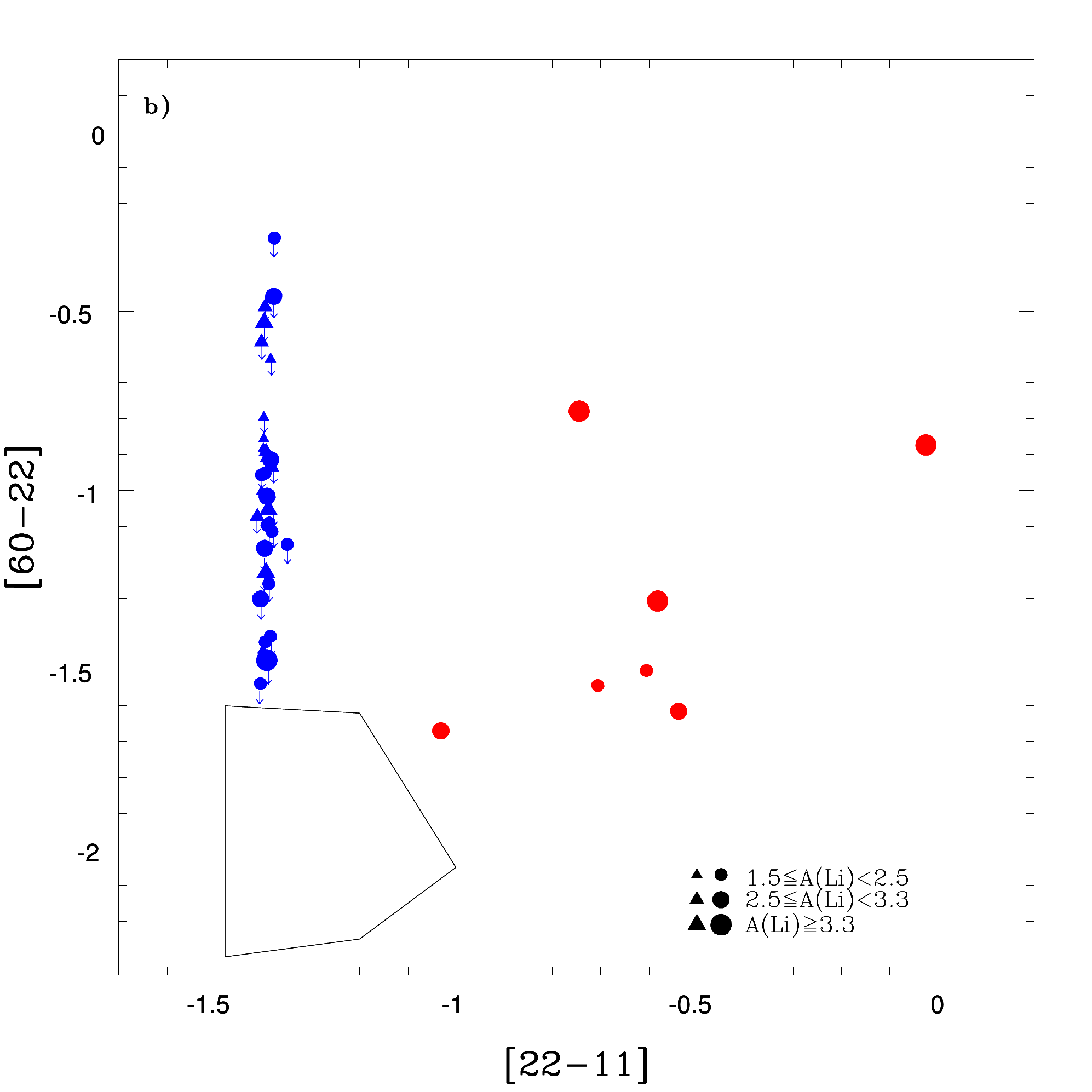}
\caption{Location of Li-rich giants in the infrared color-color diagram based on only $IRAS$ flux densities (Figure~3a) and flux densities taken 
from both $WISE$ and $IRAS$ (Figure~3b).
Li-rich giants from this survey are shown in triangles, and those from other studies are shown as circles. Red, blue, and black colors indicate good, moderate, and not good data, respectively.
The size of the symbols represents the
amount of Li, 1.5 $\leq$ A(Li) $<$ 2.5, 2.5 $\leq$ A(Li) $<$ 3.3, and
A(Li) $\geq$ 3.3. The red square denotes the Arcturus, a typical K giant with A(Li) $<$ 0.0 .}
\label{fig3}
\end{figure}

\begin{figure}[ht]
\centering
\includegraphics[scale=0.7]{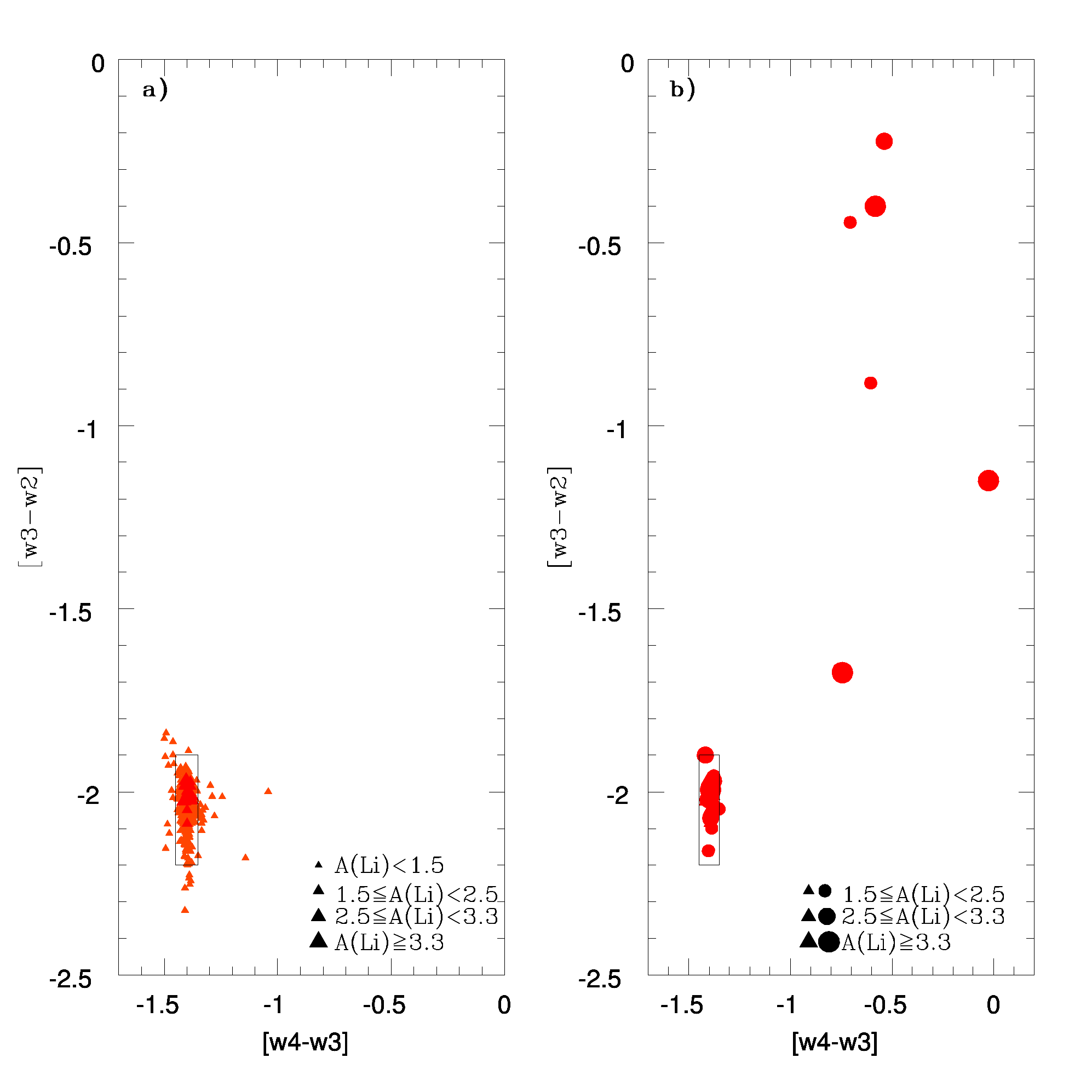}
\caption{Location of survey sample and Li-rich giants on the$WISE$ color-color diagram.
The Li-normal and Li-rich sample from this survey are shown as red triangles in the left panel. 39 Li-rich giants are shown in the right panel.}
\label{wise}
\end{figure}

\begin{figure}[ht]
\centering
 \includegraphics[scale=0.7]{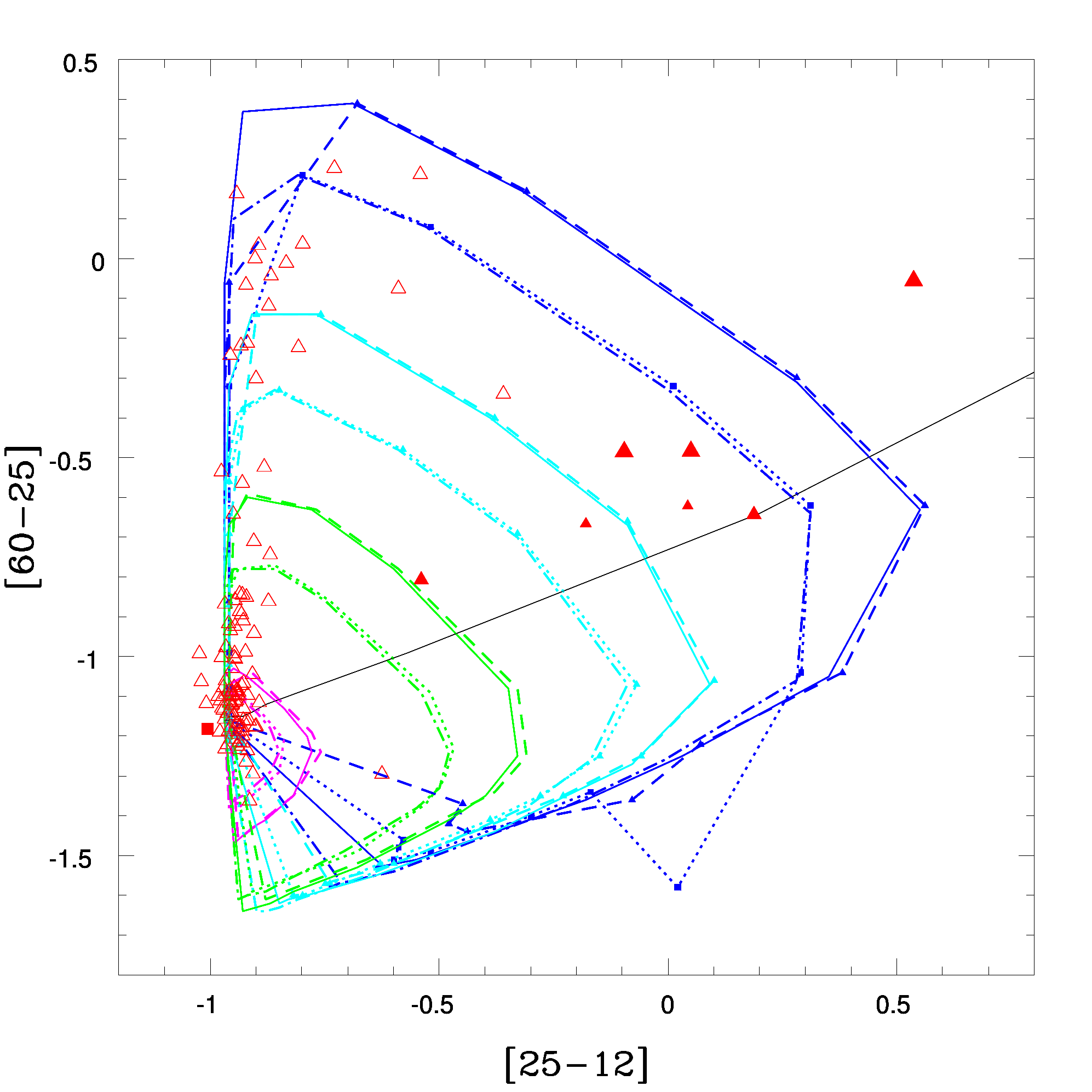}
\caption{Li-rich (filled triangles) and Li-normal (open triangles) 
giants with good quality data are superposed with dust 
shell evolutionary models.
Different colors; magenta, green, cyan, and blue represent the loops of the evolution of a dust 
shell with
mass-loss rates, 1$\times$10$^{-10}$, 1$\times$10$^{-9}$, 5$\times$10$^{-9}$, 
and 3$\times$10$^{-8}$, respectively.
Four varieties of lines in each color indicate the different photospheric temperatures and radius, 
T$_{eff}$ = 5000 K, R = 20 R$_{\sun}$, T$_{eff}$ = 4000 K, R = 20 R$_{\sun}$, T$_{eff}$ = 5000 K,
R = 10 R$_{\sun}$, T$_{eff}$ = 4000 K, R = 10 R$_{\sun}$ of the dust shell hosting stars, 
radially outward, respectively.}
\label{fig5}
\end{figure}

\begin{figure}[ht]
\centering
 \includegraphics[scale=0.7]{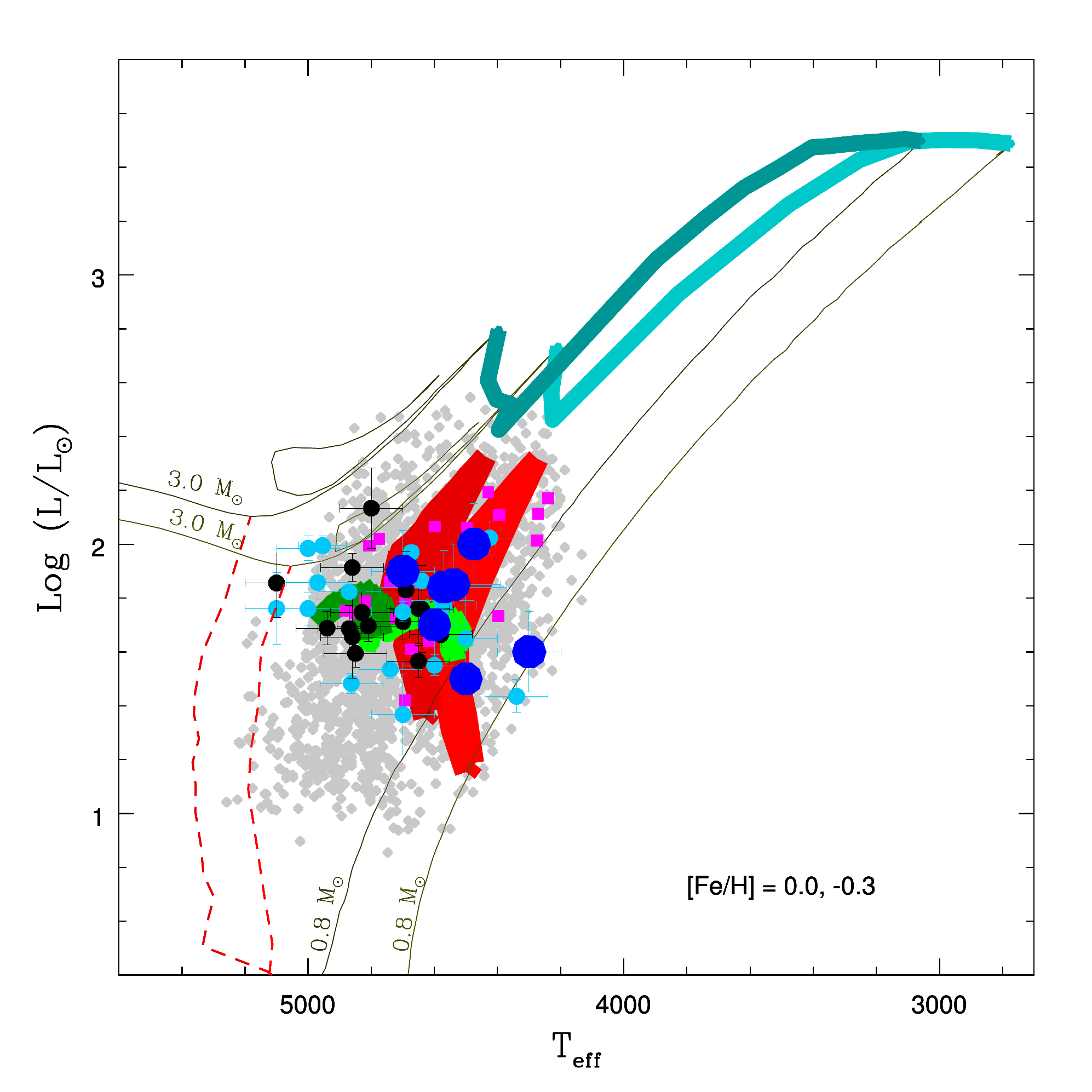}
\caption{Li-rich K giant sample (circles) and K giants with IR excess
(squares) in the HR diagram. Evolutionary phases: RGB base
(broken lines), luminosity bump (red shade), RGB tip (blue band), clump
region (green shade) are shown. The 15 Li-rich K giants taken from
\citet{bharat2011} are shown as black circles and the rest as light
blue circles. K giants with both far-IR excess and Li enhancement as shown as
big blue circles. Tracks of K giants with mass 0.8M$_{\sun}$ and 3.0M$_{\sun}$ are
shown. All the evolutionary phases are shown for two metallicity values of [Fe/H]= 0.0
and $-$0.3 .}
\label{fig6}
\end{figure}

\end{document}